\documentclass[runningheads]{llncs}

\usepackage[T1]{fontenc}
\usepackage[utf8]{inputenc}
\usepackage{amsmath,amssymb}
\usepackage{mathtools}
\usepackage{graphicx}
\usepackage{booktabs}
\usepackage{array}
\usepackage{float}
\usepackage{xcolor}
\usepackage[hidelinks]{hyperref}

\newfloat{algorithm}{t}{loa}
\floatname{algorithm}{Algorithm}
\newcounter{algln}
\newenvironment{pcode}{%
  \setcounter{algln}{0}%
  \par\vspace{2pt}\hrule\vspace{3pt}%
  \begin{list}{\footnotesize\ttfamily\stepcounter{algln}\thealgln:}%
    {\setlength{\leftmargin}{2.2em}\setlength{\itemsep}{0pt}%
     \setlength{\parsep}{0pt}\setlength{\topsep}{0pt}}%
}{%
  \end{list}\vspace{2pt}\hrule\vspace{2pt}%
}
\newcommand{\KW}[1]{\textbf{#1}}
\newcommand{\IND}{\hspace*{1.3em}}

\newcommand{\Vv}{V}
\newcommand{\E}{E}
\newcommand{\Real}{\mathbb{R}}
\begin{document}

\title{A Benchmark Generator of Realistic Logistic\\ Transport Networks with Nonlinear Edge Costs\\ and Transshipment Capacities}
\titlerunning{A Benchmark Generator of Realistic Logistic Networks}

\author{Varvara Kovaleva\inst{1} \and Alexander Ponomarenko\inst{2}}
\authorrunning{V.\,Kovaleva and A.\,Ponomarenko}

\institute{HSE University, Nizhny Novgorod, Russia\\
\email{varvara\_kovaleva03@mail.ru}
\and
Laboratory of Algorithms and Technologies for Network Analysis, HSE University,
Nizhny Novgorod, Russia\\
\email{aponomarenko@hse.ru}}

\maketitle

\begin{abstract}
Routing many commodities through a logistic network at minimum joint cost of
vehicle movements and transshipment at intermediate storages becomes
substantially harder---and substantially more realistic---once two features are
present together: edge costs that are \emph{nonlinear} in the transported
volume, because freight travels in fixed-capacity vehicles, and \emph{hard
upper bounds} on the volume reloaded at each storage. No public benchmark
captures this combination, and the locations of commercial distribution centres
are proprietary. We close this gap with a parameterised generator of synthetic
logistic networks, calibrated on real road graphs of nine European countries,
five U.S.\ states and the European part of Russia extracted from OpenStreetMap.
The generator reproduces the structural signatures of upper-level distribution
networks---near-planar topology, low vertex degree, short edges and large
diameters---by combining Zipf-distributed city sizes, a spatial-network budget
model trading edge-building cost against routing convenience,
centrality-driven storage capacities, and a doubly constrained gravity demand
model. We release 35 fully synthetic instances ($10$--$150$ storages) and 15
real-geometry instances in an open, self-describing, fully reproducible file
format. An empirical characterisation computed directly from the released files
shows that synthetic and real-geometry instances are structurally consistent
(two-sample Kolmogorov--Smirnov distances of $0.10$ on degree and edge-length
distributions, with small and quantified mean deviations), and a baseline-solver
study demonstrates that the corpus discriminates instance difficulty across
orders of magnitude. The benchmark is a faithful, reusable testbed for the
development and comparison of exact and heuristic algorithms.

\end{abstract}

% ===============================================================
\section{Introduction}
\label{sec:intro}

Planning commodity flows between distribution centres is a core problem for
logistic operators, and its realistic variants are computationally demanding.
Two features make real instances especially difficult. First, the cost of using
a road is \emph{nonlinear} in the volume carried: goods move in vehicles of
fixed capacity, so the billed quantity is the integer number of vehicles
obtained by rounding the carried volume up to a vehicle capacity. Second, every
storage has a finite \emph{transshipment capacity}, an upper bound on the volume
that can be unloaded, stored and reloaded there. The resulting model is a
capacitated multi-commodity flow problem with a step-wise vehicle cost---it is
NP-hard, and its difficulty is sensitive to both the topology and the demand
pattern of the underlying network.

Progress on solution algorithms---both exact methods and heuristics---for such
problems is hampered by the absence of suitable public benchmarks. Classical multi-commodity flow generators and libraries do
not carry node-level transshipment capacities, per-unit reload costs or
vehicle-induced nonlinear edge costs, and they rarely reflect the geographic
structure of country-scale distribution networks. Moreover, the true locations
of commercial distribution centres are proprietary. Researchers are therefore
left to test on either toy graphs or ad-hoc random graphs whose structure
differs sharply from real transport networks, which undermines the external
validity of any comparison. A generator that produces \emph{plausible} instances
at controllable scale, together with a released, citable corpus, is needed to
make comparisons of exact and heuristic algorithms meaningful and reproducible.

\paragraph{Contributions.}
This paper introduces such a generator and corpus. Specifically, we
\begin{itemize}
\item define precisely the data an instance of the problem must contain
      (Section~\ref{sec:problem});
\item analyse real road graphs from OpenStreetMap and distil, with per-region
      measurements, the structural properties a realistic generator must
      reproduce (Section~\ref{sec:real});
\item present a five-stage generation pipeline, with explicit algorithms,
      combining Zipf city sizes, a spatial-network budget model,
      centrality-driven storage capacities and a gravity demand model
      (Section~\ref{sec:pipeline});
\item describe the released benchmark, its open file format and reproducibility
      guarantees (Section~\ref{sec:corpus});
\item characterise the corpus empirically and directly from the released files,
      including two-sample Kolmogorov--Smirnov tests against the real networks
      (Section~\ref{sec:char}); and
\item demonstrate, through a baseline-solver study, that the corpus
      discriminates instance difficulty across scales
      (Section~\ref{sec:usecase}).
\end{itemize}
% ===============================================================
\section{Related Work}
\label{sec:related}

\paragraph{Flow-problem instance generators.}
Synthetic generators for network-flow problems have a long history. NETGEN
\cite{netgen} produces capacitated transportation and min-cost flow instances by
random assignment of arcs, costs and capacities, and the DIMACS implementation
challenges \cite{dimacs} standardised families of max-flow and min-cost-flow
instances. These generators target the combinatorial structure of the flow
program but are agnostic to geography: arcs and costs are sampled without
spatial embedding, so the resulting graphs do not exhibit the planar, short-edge
character of road networks, and they do not model node-level transshipment
capacities or vehicle-induced step costs. Our generator instead starts from a
realistic spatial topology and layers the logistic cost structure on top.

\paragraph{Random and spatial network models.}
Purely combinatorial models---Erd\H{o}s--R\'enyi random graphs \cite{er},
Watts--Strogatz small-world graphs \cite{ws} and Barab\'asi--Albert scale-free
graphs \cite{ba}---reproduce specific statistics (degree, clustering, path
length) but not the geometry of transport networks. Spatial network models are
closer to our needs: the Waxman model \cite{waxman} connects nodes with a
distance-decaying probability, and Gastner and Newman \cite{gastner2006} analyse
real spatial networks (road, internet and airline) and propose a construction
that trades total edge length against routing convenience. Barth\'elemy's survey
\cite{barthelemy2011} documents the near-planarity, low degree and large
diameter of road networks. We adopt the Gastner--Newman trade-off as the
topological core of our generator and calibrate its parameters against real road
graphs.

\paragraph{City sizes and travel demand.}
The rank--size distribution of city populations follows Zipf's law
\cite{gabaix1999}, which we use to seed realistic node ``masses''. Travel and
freight demand between regions is classically modelled by the gravity analogy;
the doubly constrained gravity model of Wilson \cite{wilson1967}, a staple of
transport planning \cite{ortuzar}, produces origin--destination matrices
consistent with prescribed marginals. We combine these ingredients---rarely
assembled together in a single flow-problem generator---to obtain instances that
are realistic in topology, capacity and demand simultaneously. Finally, we
follow FAIR data principles \cite{fair} in releasing a versioned, citable
corpus.

\paragraph{Heuristics for related logistic optimisation problems.}
Closely related computational work has addressed vehicle-level logistic routing
rather than upper-level flow planning. Batsyn and Ponomarenko proposed a
multi-start greedy heuristic for a real-life truck and trailer routing problem
arising in retail distribution \cite{batsyn2014ttrp}, and extended the approach
to a site-dependent variant with soft and hard time windows and split deliveries
\cite{batsyn2015heuristic}. Such methods route vehicles once the demand at each
storage is fixed; the problem targeted by the present benchmark---multi-commodity
transport between storages with nonlinear edge costs and transshipment
capacities---sits one level higher in the planning hierarchy and determines
those demands. The two problem classes are complementary, and we expect the
corpus released here to support analogous heuristic development at the
network-flow level.

% ===============================================================
\section{The Optimisation Problem an Instance Must Support}
\label{sec:problem}

To fix the data requirements we briefly state the target problem. The network is
a directed graph $G=(\Vv,\E)$ of storages $\Vv$ and roads $\E$. A set of
transportation requests (commodities) $K$ is given; each $k\in K$ has an origin
$s_k$, a destination $t_k$ and a demand $d_k\in\Real_+$. Each storage $v$ has a
per-unit reload cost $f_v\in\Real_+$ and a maximum transshipment volume
$W_v\in\Real_+$; each edge $e$ has a per-vehicle traversal cost $c_e\in\Real_+$;
and vehicles have a fixed capacity $C$. Writing $x^k_e$ for the flow of commodity
$k$ on edge $e$ and $y_e\in\mathbb{Z}_+$ for the number of vehicles on $e$, the
problem minimises movement plus reload cost,
\begin{equation}
\min\ \sum_{e\in \E} c_e\,y_e
 \;+\!\!\sum_{k\in K}\sum_{\substack{v\neq s_k,\,v\neq t_k}}\;
 \sum_{e\in\delta^-(v)} f_v\,x^k_e,
\label{eq:obj}
\end{equation}
subject to flow conservation for every commodity, the vehicle-capacity coupling
$\sum_{k} x^k_e\le C\,y_e$ for all $e\in\E$, and the storage-capacity bound
$\sum_{k}\sum_{e\in\delta^-(v)} x^k_e\le W_v$ for all transit $v$. Because the
$y_e$ are integer and the flows continuous, the model is a mixed integer linear
program (MILP), and the vehicle term makes the effective edge cost a step
function of load.

Consequently, a complete instance must specify: a weighted graph with vertex
coordinates; per-storage reload cost $f_v$ and capacity $W_v$; per-edge cost
$c_e$ (here proportional to road length); a vehicle capacity $C$; and an
origin--destination demand set $\{(s_k,t_k,d_k)\}$. The generator produces
exactly these objects, together with auxiliary data (populations, all-pairs
shortest distances) used to derive them and useful to solvers.

% ===============================================================
\section{Structural Properties of Real Networks}
\label{sec:real}

Since the locations of real distribution centres are not public, we calibrate on
city locations, populations and highways from OpenStreetMap \cite{osm}, for nine
European countries of varying size, five large U.S.\ states, and the European
part of Russia. From the raw highway graph and city list we aggregate nearby
road endpoints into nodes, select cities greedily by descending population
subject to a minimum spacing, attach each retained city to a nearby high-degree
road node (junctions are economical places for a centre), and finally simplify
the graph so that each retained centre is joined to others by edges equal to the
shortest stop-free paths. Edges are then treated as straight segments of
Euclidean length, since modelling the true road geometry would over-complicate
synthesis without materially changing the network structure.

Table~\ref{tab:real-struct} reports the resulting per-region measurements. Three
structural signatures emerge robustly and become the design targets of the
generator. First, hop diameters are large relative to size (up to $16$ for $90$
nodes); together with a right-skewed, short-edge length distribution this
indicates the \emph{opposite} of a small-world structure---routes between distant
nodes traverse many short edges. Second, the mean vertex degree is low (about
$3.6$ across the calibration graphs), well below the planar upper bound of $6$,
so the networks are essentially planar. Third, the rank--size relation of city
populations is approximately log-linear, with a Zipf exponent
$\alpha$ averaging $0.97$---consistent with the empirical range reported for
city systems \cite{gabaix1999}. The columns $\lambda$ and $coeff_{\text{budget}}$
are the spatial-network parameters (Section~\ref{sec:topo}) that best reproduce
each region's diameter, mean edge length and mean degree.

\begin{table}[t]
\centering
\caption{Structural measurements of the 15 calibration networks and the
spatial-network parameters fitted to each. $deg$: mean vertex degree;
$diam$: hop diameter; $\alpha$: fitted Zipf exponent of city sizes;
$\lambda$, $coeff_{\text{budget}}$: spatial-network model parameters
(Section~\ref{sec:topo}).}
\label{tab:real-struct}
\footnotesize
\setlength{\tabcolsep}{3.4pt}
\begin{tabular}{lrrrrr@{\hskip 1.0em}lrrrrr}
\toprule
Network & $|V|$ & $deg$ & $diam$ & $\alpha$ & $\lambda$ &
Network & $|V|$ & $deg$ & $diam$ & $\alpha$ & $\lambda$\\
\midrule
Denmark      & 11 & 2.55 & 5  & 0.91 & 0.23 & Sweden         & 45 & 3.42 & 11 & 1.04 & 0.23\\
Belgium      & 16 & 3.13 & 5  & 0.76 & 0.25 & Italy          & 46 & 3.61 & 11 & 0.76 & 0.33\\
Netherlands  & 18 & 3.67 & 5  & 0.89 & 0.03 & Germany        & 49 & 5.39 & 8  & 0.97 & 0.35\\
Hungary      & 19 & 2.74 & 7  & 1.01 & 0.23 & Texas          & 58 & 3.00 & 10 & 1.20 & 0.28\\
Ohio         & 22 & 3.73 & 6  & 0.87 & 0.03 & Spain          & 59 & 4.14 & 9  & 0.96 & 0.50\\
Georgia      & 24 & 3.75 & 7  & 0.89 & 0.08 & Eur.\ Russia   & 90 & 4.27 & 16 & 1.34 & 0.50\\
Illinois     & 28 & 3.50 & 6  & 0.88 & 0.30 & \textit{mean}  & -- & 3.57 & -- & 0.97 & 0.26\\
California   & 39 & 2.56 & 12 & 1.01 & 0.03 & UK             & 42 & 4.19 & 9  & 1.13 & 0.63\\
\bottomrule
\end{tabular}
\end{table}

% ===============================================================
\section{The Generation Pipeline}
\label{sec:pipeline}

Given a desired number of storages $k$, the generator proceeds in five stages:
city sizes, spatial layout, topology, capacities and demand. Each stage is
parameterised by distributions fitted to the calibration data, so that drawing
parameters per instance yields a diverse but realistic family.

\subsection{Stage 1 -- City sizes via Zipf's law}
A pool of cities (several times larger than $k$) is created with sizes following
Zipf's law \cite{gabaix1999}, $size_i=P/\,rank_i^{\alpha}$, where $P$ is the
largest city's size and small multiplicative noise is added for realism. From
the calibration regions (Table~\ref{tab:real-struct}) the exponent clusters near
$\alpha\approx0.97$, so the generator draws
$\alpha\sim\mathcal{N}(0.974,0.06)$, the largest population from
$(9,11)\cdot 10^4\,k$, and a city count of $k$ times a random integer in
$[8,12]$.

\subsection{Stage 2 -- Spatial layout and storage selection}
The $k$ city clusters are placed on a square map with a Gaussian blob model
(cluster standard deviation about $1/15$ of the map side), seeding the largest
city in each cluster; this imitates the real pattern of smaller towns
surrounding a large city. Storages are then selected greedily and assigned the
summed population of their nearest cities (Algorithm~\ref{alg:select}). The
minimum spacing between storages is a random value in $[80,100]$ and the map side
scales as that spacing times $k^{0.8}$.

\begin{algorithm}[t]
\caption{Greedy storage selection and population assignment.}
\label{alg:select}
\begin{pcode}
\item \KW{input:} cities $\{(p_i,pop_i)\}$, target count $k$, min.\ spacing $\rho$
\item sort cities by descending $pop_i$;\ \ $S\gets\varnothing$
\item \KW{for} each city $i$ in sorted order \KW{do}
\item \IND \KW{if} $\min_{u\in S}\lVert p_i-p_u\rVert \ge \rho$ \KW{then} $S\gets S\cup\{i\}$
\item \IND \KW{if} $|S|=k$ \KW{then break}
\item assign each city $i$ to its nearest storage $u(i)=\arg\min_{u\in S}\lVert p_i-p_u\rVert$
\item \KW{for} each storage $u\in S$:\ \ $W^{\text{pop}}_u\gets\textstyle\sum_{i:\,u(i)=u} pop_i$
\item \KW{return} storages $S$ with aggregated populations $W^{\text{pop}}$
\end{pcode}
\end{algorithm}

\subsection{Stage 3 -- Topology via a spatial-network budget model}
\label{sec:topo}
Following Gastner and Newman \cite{gastner2006}, edges are chosen to balance
construction cost against routing convenience. Construction cost is the total
edge length $\sum_{(i,j)} d_{ij}$ (Euclidean, in km), while the routing
``effective distance'' of an edge is
\begin{equation}
L_{ij}=\lambda\,d_{ij}+(1-\lambda)\cdot 1, \qquad 0\le\lambda\le1,
\end{equation}
so that $\lambda=1$ measures routes in kilometres and $\lambda=0$ in hop count.
Starting from a minimum spanning tree (which guarantees connectivity and sets the
budget reference), edges are added within a budget---a coefficient
$coeff_{\text{budget}}$ times the MST cost---so as to minimise mean pairwise
effective distance. We solve this combinatorial trade-off by simulated annealing
(Algorithm~\ref{alg:topo}). Matching the diameter, mean edge length and mean
degree of the calibration graphs yields $\lambda\sim\mathcal{N}(0.263,0.15)$ and
$coeff_{\text{budget}}\sim\mathcal{N}(2.56,0.6)$; a small budget admits only a
few long edges among many short ones, exactly the observed real pattern. Each
edge is duplicated in reverse with length differing by at most one percent, and
per-vehicle cost is $c_e=cost_{km}\cdot dist_e$ with $cost_{km}=5$.

\begin{algorithm}[t]
\caption{Budget-constrained topology by simulated annealing.}
\label{alg:topo}
\begin{pcode}
\item \KW{input:} storage positions, $\lambda$, $coeff_{\text{budget}}$
\item $T_0\gets$ MST$(V)$;\ \ $B\gets coeff_{\text{budget}}\cdot \text{cost}(T_0)$
\item $G\gets T_0$;\ \ evaluate $\Phi(G)=$ mean pairwise effective distance
\item \KW{repeat} (annealing schedule)
\item \IND propose add/remove of a candidate edge keeping $\text{cost}(G)\le B$
\item \IND accept with Metropolis probability w.r.t.\ $\Delta\Phi$
\item \KW{until} converged
\item \KW{return} connected graph $G$ within budget
\end{pcode}
\end{algorithm}

\subsection{Stage 4 -- Storage capacities and reload costs}
A storage's transshipment capacity should reflect both the demand it serves
(population) and the through-flow it attracts. The latter correlates strongly
with betweenness centrality \cite{barthelemy2011,kirkley2018}, so the capacity is
\begin{equation}
W_v = l\cdot population_v^{\gamma}\,(1+\alpha_b\,\tilde b_v),
\end{equation}
where $\tilde b_v$ is normalised betweenness, with $l=1,\gamma=0.5,\alpha_b=1.2$
and a floor $W_{\min}=100$ (capacities are scaled so at least $10\%$ fall below
the floor). Reload costs decrease with capacity, reflecting economies of scale:
$f_v=\kappa/\,W_v^{\delta}$ with $\kappa=500,\delta=1$.

\subsection{Stage 5 -- Demand via a doubly constrained gravity model}
The origin--destination matrix uses a doubly constrained gravity model
\cite{wilson1967,ortuzar},
\begin{equation}
T_{ij}=A_i\,O_i\,B_j\,D_j\,f(dist_{ij}), \qquad f(dist)=dist^{-\beta},\ \beta=1,
\end{equation}
where the marginals $O_i,D_j$ are proportional to served population (plus small
noise), $dist_{ij}$ is the shortest network distance, and the balancing factors
$A_i,B_j$ enforce $\sum_j T_{ij}=O_i$ and $\sum_i T_{ij}=D_j$, computed by the
standard iterative fitting of Algorithm~\ref{alg:gravity}. Total flow is scaled
to a few times the total storage capacity so that the capacity constraints bind.
Finally $25\%$ of requests are dropped with probability proportional to their
volume, so that not every pair exchanges goods, as in real demand matrices.

\begin{algorithm}[t]
\caption{Doubly constrained gravity balancing.}
\label{alg:gravity}
\begin{pcode}
\item \KW{input:} marginals $O,D$; deterrence $f(dist_{ij})$; tolerance $\varepsilon$
\item $A_i\gets 1,\ B_j\gets 1$ for all $i,j$
\item \KW{repeat}
\item \IND $A_i\gets \big(\sum_j B_j D_j f(dist_{ij})\big)^{-1}$ for all $i$
\item \IND $B_j\gets \big(\sum_i A_i O_i f(dist_{ij})\big)^{-1}$ for all $j$
\item \KW{until} max relative change of $A,B < \varepsilon$
\item \KW{return} $T_{ij}=A_iO_iB_jD_j f(dist_{ij})$
\end{pcode}
\end{algorithm}

% ===============================================================
\section{The Benchmark Corpus}
\label{sec:corpus}

The corpus has two parts. The \textbf{synthetic} set, produced fully by the
pipeline, contains $35$ instances: $10$ with $10$ storages, $7$ each with $20$
and $30$, $5$ with $50$, $3$ with $100$ and $3$ with $150$. The
\textbf{real-geometry} set contains $15$ instances built on the calibration
regions ($11$ to $90$ storages): the vertex coordinates and populations come from
OpenStreetMap, while edge costs, storage capacities, reload costs and demand are
generated by Stages~3--5. Together this spans $10$ to $150$ storages and $67$ to
over $16{,}000$ commodities.

Each instance is a directory of plain, self-describing files
(Table~\ref{tab:format}): an undirected weighted graph in GraphML with vertex
coordinates, a per-storage table of reload cost and capacity, a city
position/population table, the demand requests, a precomputed shortest-distance
matrix, and a parameter file recording the random seed and every generation
parameter, so that each instance regenerates exactly. The corpus and generator
are released under permissive licences (data under CC~BY~4.0, code under MIT)
following FAIR principles \cite{fair}: the instance corpus is archived on Zenodo
with a versioned DOI (\url{https://doi.org/10.5281/zenodo.20372456}), and the
generator source code is openly available on GitHub
(\url{https://github.com/Varya-K/Optimal-Logistic-Net}); an interactive web
viewer (Streamlit/Plotly) ships alongside for inspecting instances and solver
outputs.

\begin{table}[t]
\centering
\caption{File format of a benchmark instance.}
\label{tab:format}
\begin{tabular}{ll}
\toprule
File & Contents\\
\midrule
\texttt{graph.graphml}        & weighted graph; node \texttt{x,y}; edge \texttt{weight} (road length)\\
\texttt{offices.csv}          & per storage: \texttt{transfer\_price} ($f_v$), \texttt{transfer\_max} ($W_v$)\\
\texttt{pos\_pop.csv}         & city/storage \texttt{x, y, population}\\
\texttt{reqs.csv}             & demand requests \texttt{src, dst, volume} ($s_k,t_k,d_k$)\\
\texttt{distance\_matrix.csv} & all-pairs shortest road distances\\
\texttt{data.txt}             & random seed and all generation parameters\\
\bottomrule
\end{tabular}
\end{table}

% ===============================================================
\section{Empirical Characterisation}
\label{sec:char}

We characterise the corpus using statistics computed directly from the released
files, to verify that the synthetic instances reproduce real network structure.
Table~\ref{tab:syn} summarises the per-size synthetic aggregates; the figures
below pool every instance unless stated otherwise.

\begin{table}[t]
\centering
\caption{Synthetic instances aggregated by size ($n$ instances each): mean
degree, hop diameter, mean edge length, mean storage capacity $\bar W$, mean
reload cost $\bar f$, commodity count $|K|$, mean demand $\bar d$ and fitted Zipf
exponent $\alpha$. All values computed from the released files.}
\label{tab:syn}
\setlength{\tabcolsep}{6pt}
\begin{tabular}{rrrrrrrrrr}
\toprule
$|V|$ & $n$ & $deg$ & $diam$ & $dist$ & $\bar W$ & $\bar f$ & $|K|$ & $\bar d$ & $\alpha$\\
\midrule
10  & 10 & 3.68 & 3.3  & 189.5 & 266.0 & 2.63 & 67    & 153.4 & 1.00\\
20  & 7  & 4.06 & 5.3  & 261.6 & 230.4 & 2.75 & 285   & 56.7  & 0.84\\
30  & 7  & 3.25 & 7.9  & 261.9 & 199.1 & 3.00 & 652   & 36.1  & 0.79\\
50  & 5  & 3.18 & 9.4  & 304.8 & 182.8 & 3.22 & 1837  & 21.9  & 0.78\\
100 & 3  & 3.05 & 12.0 & 481.0 & 178.6 & 3.27 & 7425  & 17.2  & 0.77\\
150 & 3  & 3.15 & 18.7 & 477.0 & 177.4 & 3.30 & 16762 & 12.4  & 0.75\\
\bottomrule
\end{tabular}
\end{table}

\begin{figure}[t]
\centering
\includegraphics[width=0.92\textwidth]{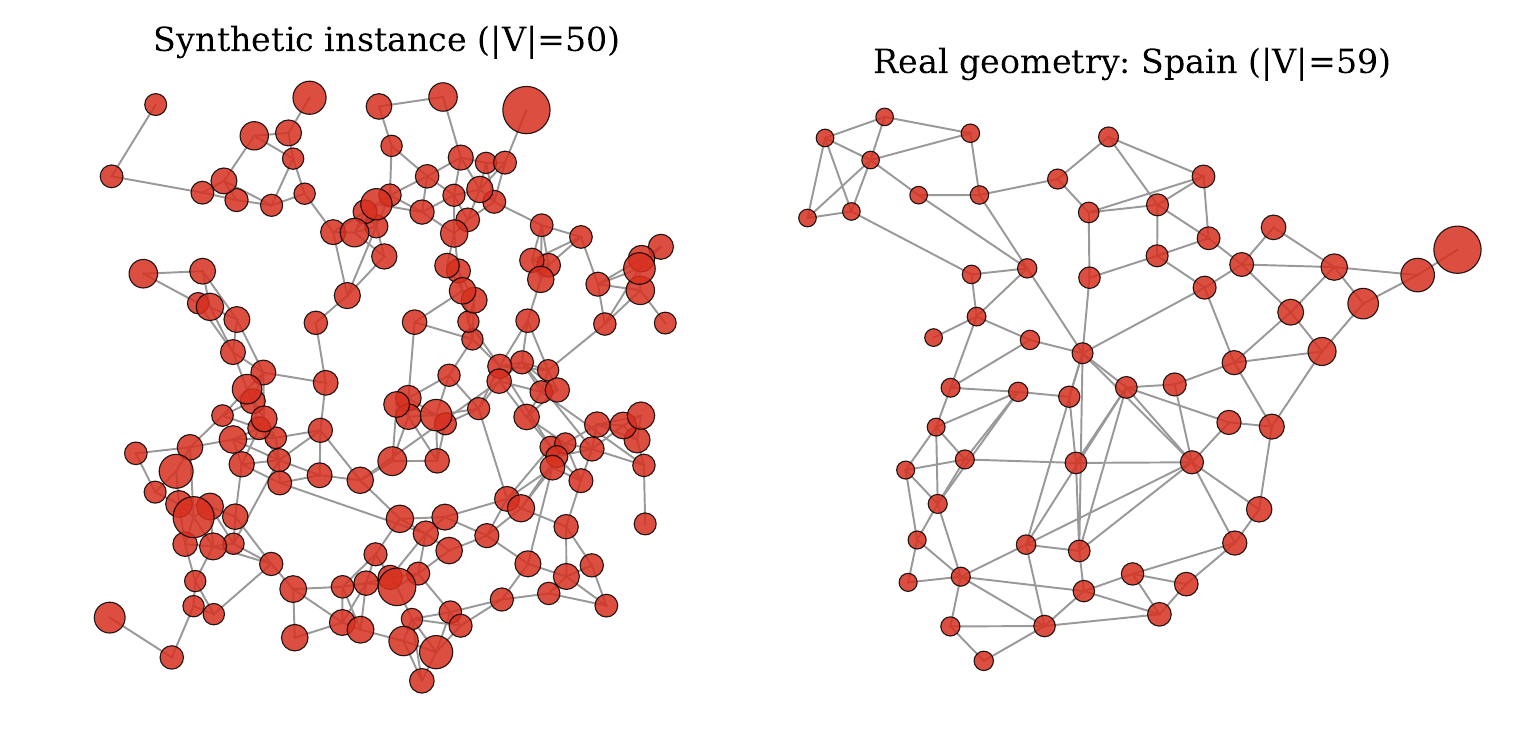}
\caption{Generated layouts. Left: a fully synthetic instance ($|V|=50$). Right: a
real-geometry instance (Spain, $|V|=59$). Node area encodes storage capacity.}
\label{fig:layouts}
\end{figure}

\begin{figure}[t]
\centering
\includegraphics[width=\textwidth]{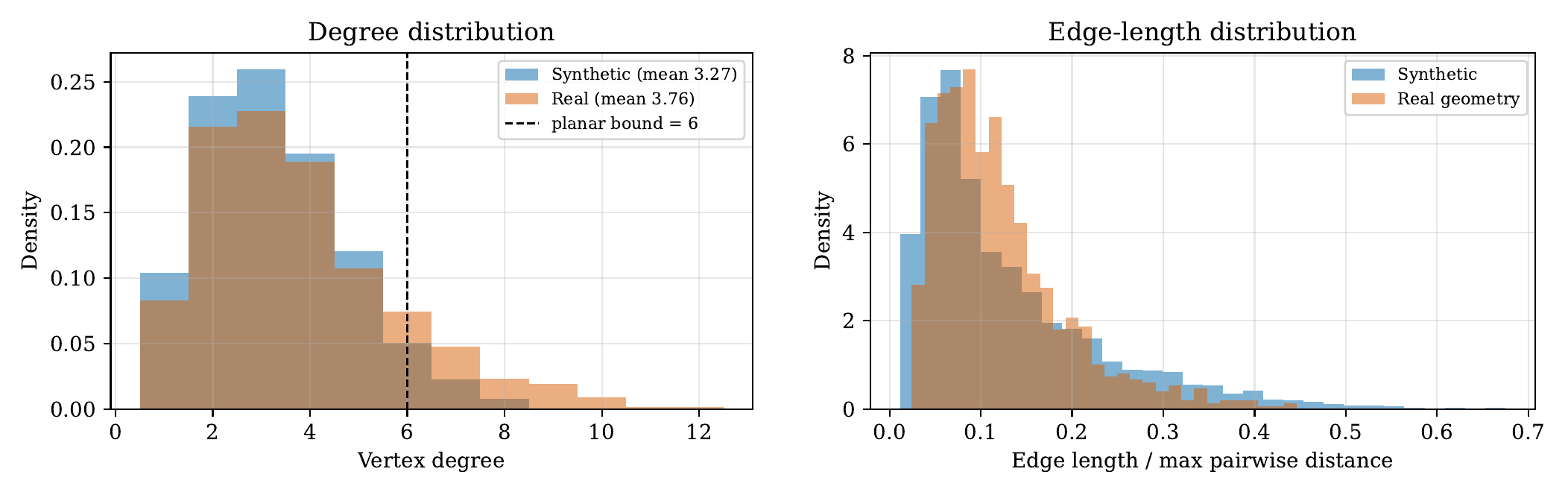}
\caption{Pooled topological distributions over all instances. Left: vertex
degree (both populations stay below the planar bound $6$). Right: normalised
edge length (short edges dominate). Synthetic and real-geometry instances are
closely aligned.}
\label{fig:topodist}
\end{figure}

\begin{figure}[t]
\centering
\includegraphics[width=\textwidth]{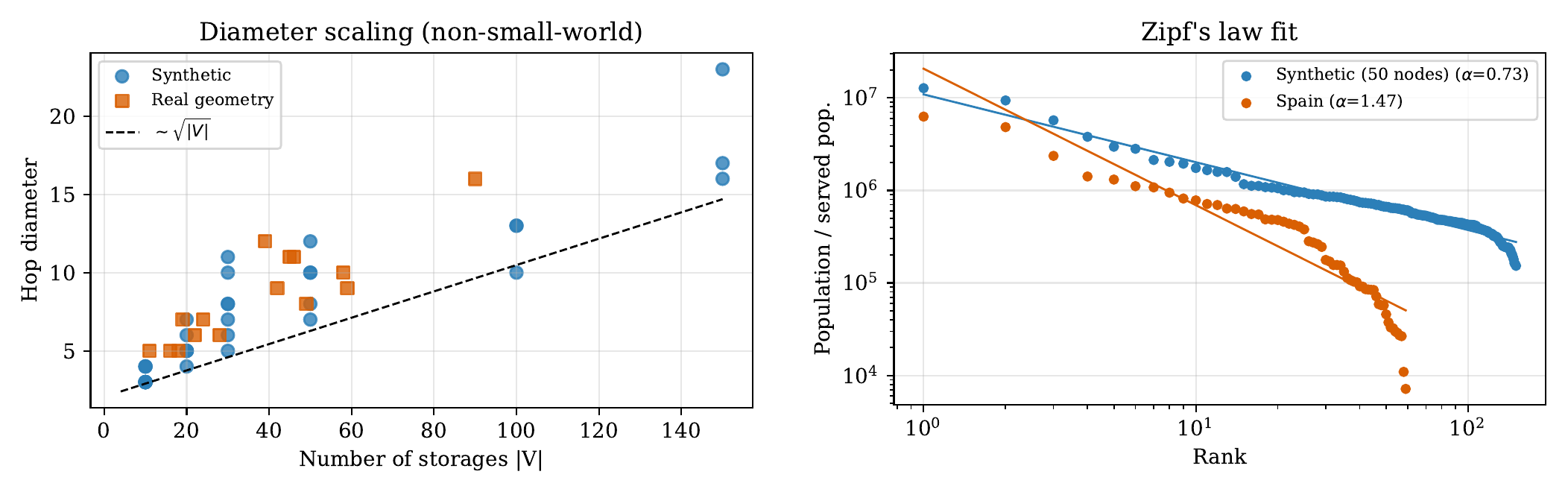}
\caption{Left: hop diameter grows faster than $\sqrt{|V|}$ for both populations
(networks are not small-world). Right: the rank--size relation of storage
populations is approximately log-linear (Zipf).}
\label{fig:diamzipf}
\end{figure}

\paragraph{Topology.}
Figure~\ref{fig:layouts} shows a synthetic instance beside the Spain network:
both display the same near-planar, locally clustered structure with mostly short
edges. Figure~\ref{fig:topodist} pools all instances: the degree distribution is
concentrated below the planar bound of $6$, and the edge-length distribution
(normalised by each graph's largest pairwise distance) is right-skewed and
dominated by short edges, in both populations. Figure~\ref{fig:diamzipf} confirms
that hop diameter grows faster than the $\sqrt{|V|}$ reference (non-small-world),
with synthetic and real points overlapping, and that the storage population
rank--size relation is approximately log-linear, as Zipf's law prescribes.

\paragraph{Quantitative agreement.}
To test agreement formally we ran two-sample Kolmogorov--Smirnov (KS) tests
between the pooled synthetic and real-geometry distributions. The KS distances
are small---$D=0.095$ for vertex degree ($n_{\text{syn}}=1450$,
$n_{\text{real}}=566$) and $D=0.103$ for normalised edge length
($n_{\text{syn}}=2373$, $n_{\text{real}}=1064$)---confirming that the
distributions are close in shape. Both differences are nonetheless statistically
significant ($p<0.01$), which at these large pooled sample sizes reflects modest
but real deviations rather than a poor fit: mean degree is $3.27$
($95\%$ CI $[3.20,3.35]$) for synthetic versus $3.76$ ($[3.60,3.92]$) for real,
and mean normalised edge length is $0.130$ ($[0.126,0.134]$) versus $0.121$
($[0.117,0.125]$). In other words, the synthetic networks are slightly sparser
and have marginally longer normalised edges, but reproduce the real
distributions' location, shape and support faithfully. The same qualitative
trends across size (Table~\ref{tab:syn}) hold for the real networks
(Table~\ref{tab:real-struct}), supporting the use of the synthetic set as a
controllable proxy for real geography.

% ===============================================================
\section{Use Case: Difficulty and Solver Discrimination}
\label{sec:usecase}

A benchmark is only useful if its instances vary in difficulty. We quantify
difficulty with three standard, solver-agnostic quantities, all expressed
relative to the linear-relaxation optimum (LP), which is a valid lower bound on
the true optimum. The first is the LP value itself (ratio $1$ by definition). The
second is the trivial feasible solution obtained by rounding the relaxation up to
integer vehicle counts (Rounded LP); the ratio $\text{RoundedLP}/\text{LP}\ge1$
brackets the optimality gap and needs no solver beyond a linear program. The
third is the best feasible solution returned by a general-purpose exact MILP
solver, HiGHS \cite{highs}, under a wall-clock budget proportional to instance
size; the ratio $\text{MILP}/\text{LP}\ge1$ measures how far a standard exact
method gets within a fixed budget. None of these quantities depends on a bespoke
algorithm, so the difficulty they reveal is a property of the instances, not of
any particular method.

\begin{figure}[t]
\centering
\includegraphics[width=\textwidth]{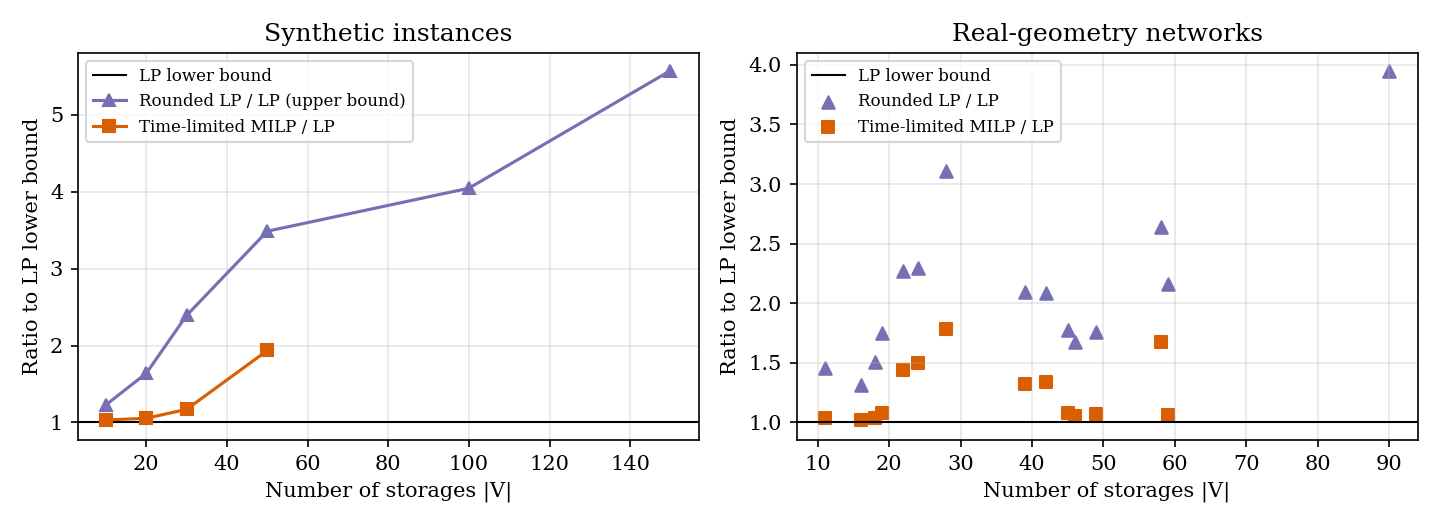}
\caption{Instance difficulty as ratios to the LP lower bound versus network size.
Left: synthetic instances (means per size). Right: real-geometry networks. The
trivial upper bound diverges from the lower bound and the time-limited exact
solver becomes infeasible as size grows, demonstrating that the corpus
discriminates difficulty without reference to any bespoke solver.}
\label{fig:difficulty}
\end{figure}

Table~\ref{tab:usecase} and Figure~\ref{fig:difficulty} show two clear trends.
First, the relaxation gap widens monotonically with size: the trivial upper bound
sits $1.23\times$ above the LP lower bound at $|V|=10$ but $5.6\times$ above it at
$|V|=150$, so the interval that any solver must close grows by roughly an order of
magnitude across the corpus. Second, the exact solver degrades and then fails
exactly where the gap widens: $\text{MILP}/\text{LP}$ rises from $1.03$ at
$|V|=10$ to $1.94$ at $|V|=50$, and for $|V|\ge100$ HiGHS returns no feasible
solution at all within the budget. The corpus therefore spans the full range from
instances an exact solver closes to near-optimality to instances on which it
fails outright, which is precisely the discrimination a benchmark must provide.
The real-geometry instances (Fig.~\ref{fig:difficulty}, right) reproduce these
trends: the upper-bound ratio ranges from $1.31$ on the smallest networks to
$3.95$ on European Russia ($|V|=90$), and the exact solver is feasible only up to
$|V|=59$. That the same hardness signature appears on networks built from real
geography confirms that difficulty, and not only structure, transfers between the
synthetic and real-geometry sets.

\begin{table}[t]
\centering
\caption{Instance difficulty on the synthetic set, aggregated by size and
expressed as ratios to the LP lower bound (mean over the $n$ instances with
reference solutions). RoundedLP/LP brackets the optimality gap; MILP/LP is the
time-limited exact solver. ``--'' marks sizes where HiGHS returned no feasible
solution within the budget.}
\label{tab:usecase}
\setlength{\tabcolsep}{8pt}
\begin{tabular}{rrrr}
\toprule
$|V|$ & $n$ & Rounded LP / LP & Exact MILP / LP\\
\midrule
10  & 10 & 1.229 & 1.032\\
20  & 7  & 1.640 & 1.055\\
30  & 7  & 2.391 & 1.171\\
50  & 5  & 3.485 & 1.935\\
100 & 3  & 4.044 & --\\
150 & 1  & 5.572 & --\\
\bottomrule
\end{tabular}
\end{table}

% ===============================================================
\section{Limitations and Future Work}
\label{sec:limits}

The generator makes deliberate simplifications. Edges are straight segments, so
true road sinuosity is not modelled (it changes lengths by a near-constant factor
and does not affect topology). Demand is static and symmetricised; time-dependent
or stochastic demand is left for future work. Capacities derive from population
and betweenness only; richer drivers (warehouse type, modal mix) could be added.
The KS analysis shows the synthetic networks are slightly sparser than the real
ones, suggesting the budget prior could be widened for larger $k$. Each of these
is a parameter or stage the modular pipeline can extend without redesign, and the
released code is structured to make such extensions straightforward.

% ===============================================================
\section{Conclusion}
\label{sec:conclusion}

We presented a generator and an open, reproducible corpus for logistic
network-flow problems with nonlinear, vehicle-induced edge costs and hard storage
transshipment capacities---a problem class for which no realistic public
benchmark previously existed. By calibrating on real road graphs and combining
Zipf city sizes, a spatial-network budget model, centrality-driven capacities and
a gravity demand model, the generator reproduces the near-planar topology, low
degree, short edges and large diameters of real upper-level distribution
networks. An empirical characterisation computed directly from the released
files shows that the $35$ synthetic and $15$ real-geometry instances are
structurally consistent (KS distances $\approx0.1$ with small, quantified mean
deviations), and a baseline-solver study demonstrates that the corpus
discriminates instance difficulty across orders of magnitude. The benchmark,
generator and viewer are released together so that comparisons of exact and
heuristic algorithms on this problem can be reproducible and externally valid.

\bigskip
\noindent\textbf{Data availability.} The instance corpus is openly available
under CC~BY~4.0 and archived on Zenodo with a versioned DOI,
\url{https://doi.org/10.5281/zenodo.20372456}. The generator and the interactive
viewer are released under the MIT licence in the development repository on
GitHub, \url{https://github.com/Varya-K/Optimal-Logistic-Net}.

\appendix
\section{Per-Region Instance Statistics}
\label{app:real}
Table~\ref{tab:real-inst} lists the logistic parameters generated for the
real-geometry instances: total storage capacity $W_{\text{sum}}$, mean reload
cost $\bar f$, number of commodities $|K|$ and total demand $d_{\text{sum}}$.
All values are computed from the released files and complement the structural
measurements of Table~\ref{tab:real-struct}.

\begin{table}[h]
\centering
\caption{Logistic parameters of the 15 real-geometry instances.}
\label{tab:real-inst}
\footnotesize
\setlength{\tabcolsep}{3.6pt}
\begin{tabular}{lrrrr@{\hskip 0.8em}lrrrr}
\toprule
Network & $W_{\text{sum}}$ & $\bar f$ & $|K|$ & $d_{\text{sum}}$ &
Network & $W_{\text{sum}}$ & $\bar f$ & $|K|$ & $d_{\text{sum}}$\\
\midrule
Denmark     & 3294  & 2.35 & 82   & 8323   & Sweden       & 21373 & 1.83 & 1485 & 85688\\
Belgium     & 5703  & 2.11 & 180  & 15717  & Italy        & 14288 & 2.40 & 1552 & 61953\\
Netherlands & 4694  & 2.73 & 229  & 11605  & Germany      & 14276 & 2.54 & 1764 & 43705\\
Hungary     & 4566  & 3.26 & 256  & 20566  & Texas        & 27767 & 2.31 & 2479 & 110540\\
Ohio        & 7371  & 2.58 & 346  & 22994  & Spain        & 19789 & 2.38 & 2566 & 77666\\
Georgia     & 7720  & 2.57 & 414  & 34832  & Eur.\ Russia & 42492 & 2.03 & 6007 & 132570\\
Illinois    & 10292 & 2.53 & 567  & 24337  & UK           & 24109 & 1.75 & 1291 & 77119\\
California  & 36544 & 1.48 & 1111 & 114359 & & & & &\\
\bottomrule
\end{tabular}
\end{table}

\subsubsection*{Acknowledgments.}
The study was implemented in the framework of the Basic Research Program at
HSE University (HSE-BR-2025-080).

\end{document}